\def\year{2021}\relax
\documentclass[letterpaper]{article} 
\usepackage{aaai21}  
\usepackage{times}  
\usepackage{helvet} 
\usepackage{courier}  
\usepackage[hyphens]{url}  
\usepackage{graphicx} 
\usepackage{booktabs}       
\usepackage{amsfonts}       
\usepackage{nicefrac}       
\usepackage{microtype}      
\usepackage{amsmath}
\usepackage{amssymb}
\usepackage{subcaption}
\usepackage{gensymb}
\usepackage{float}
\usepackage{mathtools}
\usepackage{multirow}
\usepackage{tabularx,ragged2e}
\usepackage{enumitem}
\usepackage{bm}
\usepackage{diagbox, xcolor}
\usepackage[switch]{lineno}
\usepackage[ruled,vlined]{algorithm2e}
\usepackage{footmisc}

\usepackage{natbib}  
\usepackage{caption} 
\title{XraySyn: Realistic View Synthesis From a Single Radiograph Through CT Priors}
\author{Cheng Peng\textsuperscript{1}\footnotemark[1] \hspace{2mm} Haofu Liao\textsuperscript{2}\footnotemark[1]
 \hspace{2mm} Gina Wong \textsuperscript{1} \hspace{2mm} Jiebo Luo\textsuperscript{2} \hspace{2mm}  Shaohua Kevin Zhou\textsuperscript{3,4} \hspace{2mm} Rama Chellappa\textsuperscript{1} \\
\textsuperscript{1}Johns Hopkins University \hspace{2mm}
\textsuperscript{2}University of Rochester \hspace{2mm}\\
\textsuperscript{3}Chinese Academy of Sciences \hspace{2mm}
\textsuperscript{4}Peng Cheng Laboratory, Shenzhen}
\affiliations{}
\begin{document}
\maketitle
\begin{abstract}
\footnotetext[1]{Equal contribution. Cheng proposed and implemented the XraySyn framework for view synthesis and bone suppression, labeled data, and performed relevant experiments. Haofu initiated the project, proposed and implemented the differentiable forward/backprojector module, and provided valuable discussions.}
A radiograph visualizes the internal anatomy of a patient through the use of X-ray, which projects 3D information onto a 2D plane. Hence, radiograph analysis naturally requires physicians to relate the prior about 3D human anatomy to 2D radiographs. Synthesizing novel radiographic views in a small range can assist physicians in interpreting anatomy more reliably; however, radiograph view synthesis is heavily ill-posed, lacking in paired data, and lacking in differentiable operations to leverage learning-based approaches. To address these problems, we use Computed Tomography (CT) for radiograph simulation and design a \textbf{differentiable projection} algorithm, which enables us to achieve geometrically consistent transformations between the radiography and CT domains. Our method, XraySyn, can synthesize novel views on \textbf{real radiographs} through a combination of realistic simulation and finetuning on real radiographs. To the best of our knowledge, this is \textbf{the first work} on radiograph view synthesis. We show that by gaining an understanding of radiography in 3D space, our method can be applied to radiograph bone extraction and suppression without groundtruth bone labels.


\end{abstract}
\section{Introduction}

Radiography, widely used for visualizing the internal human anatomy, applies high-energy radiation, or X-ray, to pass through the body, and measures the remaining radiation energy on a planar detector. Since different organs attenuate X-ray to various degrees, the detected energy is visualized as a 2D image or a radiograph, that reveals the internal structure of the body and provides valuable diagnostic information.



\begin{figure}[!htb]
    \setlength{\abovecaptionskip}{3pt}
    \setlength{\tabcolsep}{2pt}
    \begin{tabular}[b]{cc}
        \begin{subfigure}[b]{.48\linewidth}
            \includegraphics[width=\textwidth,height=\textwidth]{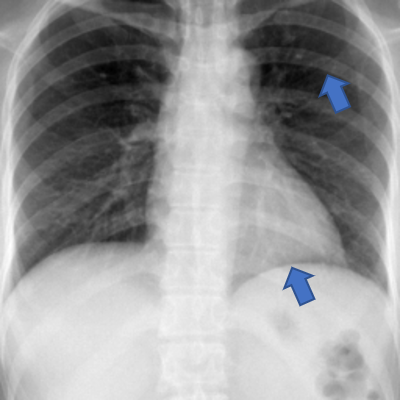}
            \caption{Input Radiograph}
            \label{input}
        \end{subfigure} &
        \begin{subfigure}[b]{.48\linewidth}
            \includegraphics[width=\textwidth,height=1\textwidth]{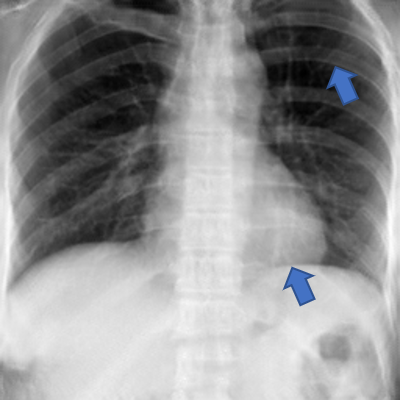}
            \caption{Synthesized Novel View}
            \label{novel_view}
        \end{subfigure} \\
        \begin{subfigure}[b]{.48\linewidth}
            \includegraphics[width=\textwidth,height=1\textwidth]{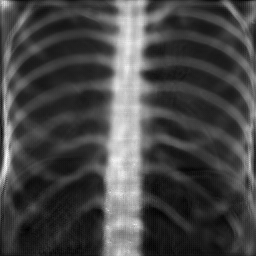}
            \caption{Bone Estimation on Input}
            \label{bone_est}
        \end{subfigure} &
        \begin{subfigure}[b]{.48\linewidth}
            \includegraphics[width=\textwidth,height=1\textwidth]{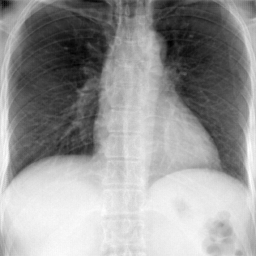}
            \caption{Bone Suppression on Input}
            \label{bone_suppress}
        \end{subfigure} \\
    \end{tabular}
    \caption{From (a) a \emph{real} radiograph, XraySyn synthesizes (b) a radiograph of novel view. As the view point rotate clockwise in azimuth angle, observe that the heart and the rib bones, as the blue arrows indicate, change accordingly.
Additionally, XraySyn obtains (c) the bone structure across all views and can be used to perform (d) bone suppression. Both synthesized views and bone estimation are generated without direct supervision. }
    \label{fig:intro}
    \vspace{-1.5em}
\end{figure}
Radiography is fast and economical; however, high energy radiation can cause adverse health effects. Conventionally only a single, frontal view radiograph is acquired per session, e.g. for chest radiography. While physicians can intuitively relate the different organs on a 2D radiograph in 3D space, such intuition is implicit and varies in accuracy.  As such, a radiograph view synthesis algorithm can help provide additional information to assist in understanding a patient's internal structure. 
The ability to understand radiographs in a 3D context is also essential beyond providing more visual information. For example, since every pixel on a radiograph represents an X-ray traversing in 3D, it is principally ambiguous to label a pixel as a specific anatomical structure, as X-rays inevitably pass through multiple structures. By exploring 3D context, we can disentangle a pixel into values that represent different structures, and lead to improved analysis algorithms. Particularly, bone extraction on \emph{real} radiographs is difficult, but has been used in fracture analysis \cite{DBLP:journals/corr/abs-2007-01464}, lesion detection \cite{li2020high}, etc. In this work, we tackle radiograph understanding in 3D, specifically through the tasks of novel view synthesis and bone extraction. 

Transforming 2D images to 3D objects is by nature ill-posed. In the natural image domain, deep learning-based methods have shown impressive results in addressing such a problem in a data-driven way. Radiograph view synthesis, however, poses several unique challenges. Firstly, there is no multi-view dataset for radiographs due to privacy and radiation concerns, which prohibits the use of supervised learning. Secondly, there lacks a \emph{differentiable} algorithm that ensures geometrically consistent transformations between radiographs and the 3D space. Lastly, unlike visible light, X-rays can penetrate objects, therefore to inverse the X-ray projection one should take into account both the surface and the internal information in 3D, making the problem even more ill-posed than for natural images and thus poses challenges to unsupervised methods. 

While inverse graphics is a daunting task to be solved directly on real radiographs, some attempts have been made to address it through Digitally Reconstructed Radiographs (DRRs) \cite{DBLP:conf/cvpr/YingGMWWZ19,DBLP:journals/cgf/HenzlerRRR18}. DRRs  \cite{doi:10.1259/bjr/30125639} are simulated radiographs from Computed Tomography (CT) volumes, which are abundantly available. This approach addresses the data scarcity for learning a 2D-3D transformation; however, there remains significant differences between real radiographs and DRRs. The generation of DRRs also cannot be incorporated into a learning-based algorithm, and either is done off-line \cite{DBLP:journals/cgf/HenzlerRRR18} or has severe limitations in available view angles \cite{DBLP:conf/cvpr/YingGMWWZ19}.


In this work, we introduce a novel, two-stage algorithm called \emph{XraySyn} to estimate the 3D context from a radiograph and uses it for novel view synthesis and bone extraction. The first stage of XraySyn, called 3D PriorNet (3DPN), incorporates a pair of differentiable backprojection and forward projection operators to learn the radiograph-to-CT transformation under simulated setting. These operators ensure the transformations between radiograph and CT to be geometrically consistent, therefore significantly reducing the complexity of learning. We further incorporate the differentiable forward projector into a modified DeepDRR \cite{DBLP:conf/miccai/UnberathZLBFAN18}, which simulates realistic radiographs, to minimize the domain gap between DRRs and real radiographs. 
The second stage of XraySyn, called 2D RefineNet (2DRN), further enhances the projected radiograph from its estimated 3D CT. By using a Generative Adversarial Network (GAN) and residual connections, 2DRN produces high quality radiographic views and their respective bone structure. 
In summary, our contributions can be described in four parts:
\begin{enumerate}
    \item We propose a differentiable forward projection operator and incorporate it within a modified DeepDRR, forming a pipeline called CT2Xray that simulates realistic radiographs from CT, propagates gradients, and runs fast.
    \item We propose a 3D PriorNet (3DPN), which incorporates CT2Xray and generates the 3D context from a single radiograph through learning from the paired relationship between simulated radiographs and their CT volumes.
    \item We propose a 2D RefineNet (2DRN), which refines on the 2D radiographs projected from 3DPN's output. By leveraging the availability of CT labels and the CT2Xray pipeline, 2D RefineNet can synthesize not only novel radiograph views, but also the corresponding bone structure.
    \item We examine XraySyn, comprising 3DPN and 2DRN, on real radiographs and find the performance of view synthesis and bone extraction visually accurate, despite the lack of direct supervision in the radiograph domain.
\end{enumerate}








\section{Related work}

\subsubsection{View synthesis from a single image }

There is a long history of research in natural image view synthesis. For relevancy and brevity, we focus on recent advances in view synthesis based on a single image and with the use of CNN. One approach to tackling such a task is to generate the new view in an image-to-image fashion. Some methods \cite{DBLP:conf/nips/ChenCDHSSA16,DBLP:conf/nips/KulkarniWKT15} propose to generate a disentangled space where the image can be projected to and modified from to synthesize new views, while others \cite{DBLP:conf/cvpr/ParkYYCB17,DBLP:conf/eccv/SunHLZL18,DBLP:conf/eccv/TatarchenkoDB16,DBLP:conf/eccv/ZhouTSME16} rely on GANs to generate the information that is occluded from the original view. In general, the image-to-image approach is based on sufficient pixel correspondence between views, which provide understanding for recovery in either the image space or latent space. Such pixel correspondence is much weaker between X-ray views. In spirit, our method is more similar to the 3D shape generation approaches \cite{DBLP:conf/eccv/GirdharFRG16,DBLP:conf/eccv/ChoyXGCS16,DBLP:journals/corr/abs-1911-13225,DBLP:conf/nips/XuWCMN19,DBLP:conf/iccv/Gkioxari0M19,DBLP:journals/corr/abs-1802-05384} that concern the generation of 3D surfaces, which are less ill-posed than generating 3D volumes. 

\subsubsection{Radiograph simulation and transformation to CT}
Due to the lack of multi-view radiographs and the difficulties in correctly labelling them, data-driven methods that require large number of radiographs often turn to CT-based radiograph simulations. While Monte-Carlo (MC) methods based on imaging physics \cite{badal2009accelerating,sisniega2013monte,schneider2000correlation} can lead to highly realistic radiograph simulations, they are time-consuming and not scalable. Many works \cite{li2020high,DBLP:conf/miccai/GozesG18,DBLP:conf/cvpr/YingGMWWZ19,DBLP:conf/miccai/AlbarqouniFN17,8363572} use DRRs, which are less realistic but computationally inexpensive radiograph simulations, to perform tasks such as bone enhancement, bone suppression, disease identification, CT reconstruction, etc. In particular, Ying \cite{DBLP:conf/cvpr/YingGMWWZ19} addresses the discrepancy between DRRs and real radiographs by training an additional domain adaptation network. Henzler \cite{DBLP:journals/cgf/HenzlerRRR18} uses real cranial X-ray images acquired in a controlled setting to recover the 3D bone structure. As clinical evidence supports that bone suppression on radiograph can improve diagnostic accuracy \cite{laskey1996dual}, Li \cite{li2020high} proposes to achieve bone suppression by learning a bone segmentation network based on DRRs, and apply the network on real radiographs with handcrafted post-processes. Recently, DeepDRR \cite{DBLP:conf/miccai/UnberathZLBFAN18} is proposed to model DRR generation more accurately by replicating similar procedures from the MC simulation counterpart, and shows that CNN models trained on such simulations are able to generalize better on real radiographs.

\section{Method}
The main goal of this work is to synthesis novel views from a frontal view radiograph, which requires a degree of 3D knowledge. As multi-view dataset is not readily available for real radiographs, our proposed method, XraySyn, composes of two stages. The first stage learns to estimate 3D knowledge under a simulated setting by using CT volumes. The second stage transfers such learning to generate real radiographs. 

The challenge underlying this approach is how to best address the transformation from simulated radiographs to CT volumes, while ensuring the input radiograph can be reproduced from such volume. We first explain the proposed operators that enable learning such a transformation, CT2Xray and Single Image Backprojector. We then introduce XraySyn, which incorporates the two operators for a radiograph-to-CT-to-radiograph algorithm.


\begin{figure*}[htb!]
\centering
\includegraphics[width=\textwidth]{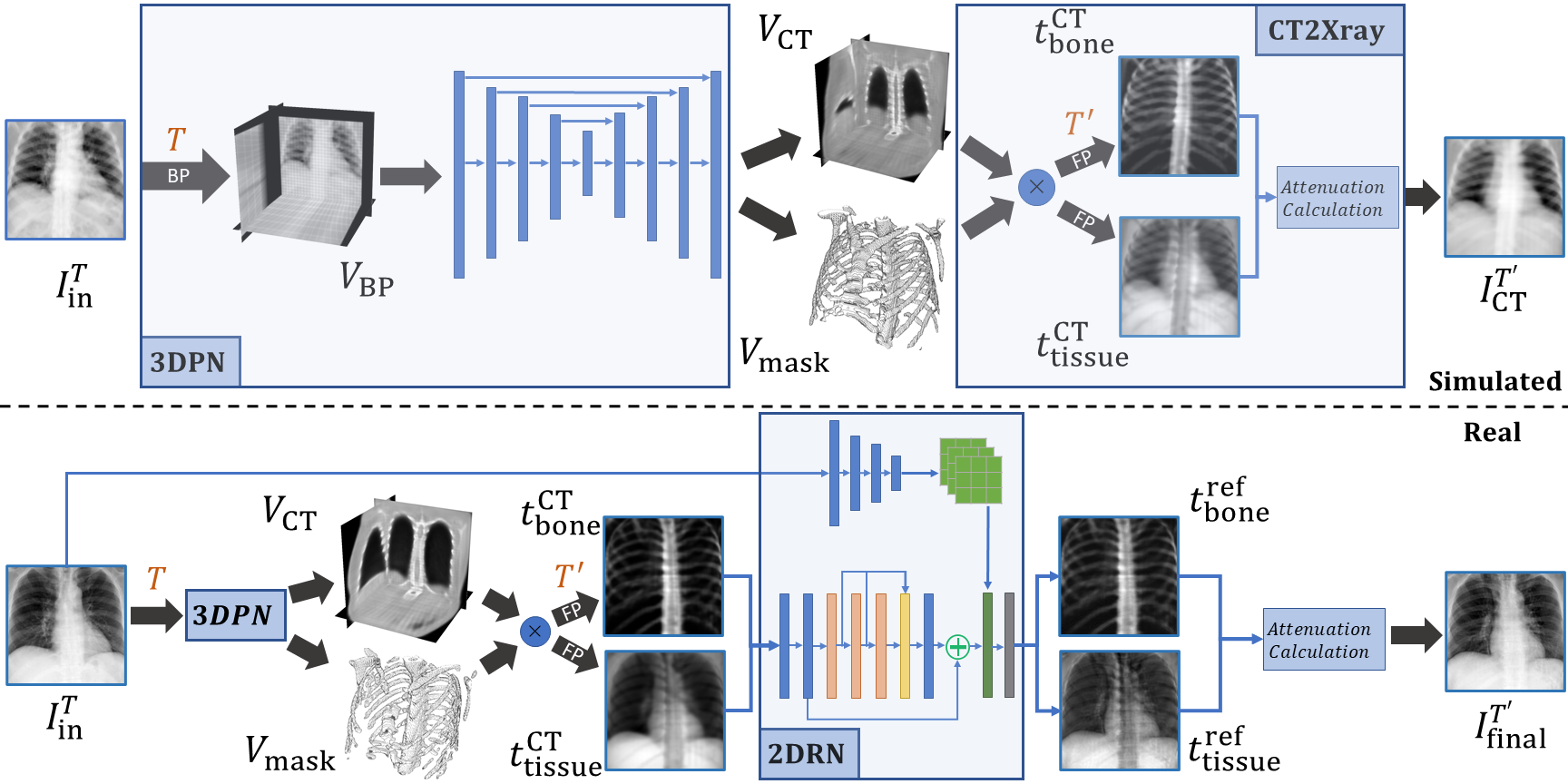}
\caption{The proposed two-stage network structure of XraySyn.
In the first stage (top), \emph{simulated} radiographs are backprojected (BP) from view $T$, and refined through 3DPN to obtain their CT and bone mask estimates, $V_{\textrm{CT}}$ and $V_{\textrm{mask}}$. Through CT2Xray, $V_{\textrm{CT}}$ and $V_{\textrm{mask}}$ are forward projected (FP) from view $T^{'}$ to calculate the tissue and bone content $t^{\textrm{CT}}_{\textrm{bone}}$ and $t^{\textrm{CT}}_{\textrm{tissue}}$, which are used to simulate the novel view radiograph. In the second stage (bottom), $t^{\textrm{CT}}_{\textrm{bone}}$ and $t^{\textrm{CT}}_{\textrm{tissue}}$ are generated from \emph{real} radiographs through a trained 3DPN, refined through 2DRN, and used to generate the novel view.} 
\label{fig:xraysyn_net}
\vspace{-1.5em}
\end{figure*}
\subsection{CT2Xray}
CT2Xray, as shown in Fig. \ref{fig:xraysyn_net}, has two parts: (i) the differentiable forward projection; (ii) attenuation-based radiograph simulation, which along with the first part forms CT2Xray and transforms a CT volume into a realistic radiograph. To the best of our knowledge, CT2Xray is \textit{the first algorithm} that can generate realistic radiographs from CT volumes with gradient propagation along arbitrary viewpoint.

\textbf{Differentiable forward projector (FP)}.
Let $V_{\textrm{CT}}$ denote a CT image, FP generates a 2D projection of $V_{\textrm{CT}}$ by:
\begin{align} \label{eq:drr}
\begin{aligned}
    I(x) = \textrm{FP}(V_{\textrm{CT}}, T) = \int 
    V_{\textrm{CT}}(T^{-1} p) d\bm{l}(x) \\
     ~\approx \sum_{p \in \bm{l}(x)} V_{\textrm{CT}}(T^{-1} p) \Delta p,
\end{aligned}
\end{align}
where $T$ is a homogeneous matrix that controls the rotation and translation of the view point and $\bm{l}(x)$ is a line segment connecting the simulated x-ray source and detector at $x$. For backpropagation, the gradients of $I$ wrt $V_{\textrm{CT}}$ can be written as
\begin{equation}\label{eq:grad}
 \dfrac{\partial I(x)}{\partial V_{\textrm{CT}}(y)} =
 \begin{cases}
  \Delta p, &\text{if} ~~ Ty \in \bm{l}(x),\\
     0, &\text{otherwise. }
 \end{cases}
\end{equation}
Equations for (\ref{eq:drr}) and (\ref{eq:grad}) can be implemented through massive parallelism with GPUs, where every line integral over the volume is a standalone operation; as such this implementation can be used in online training.

\textbf{CT2Xray}.\label{sec:ct2xray}
Forward projecting CT volumes generates DRRs, which are poor simulations of X-ray images due to the inaccurate assumption that different tissues attenuate the X-ray similarly. DeepDRR \cite{DBLP:conf/miccai/UnberathZLBFAN18} produces a better simulation by avoiding this assumption, but it is not differentiable and hence not amenable for end-to-end learning. We contribute a \textit{differentiable 
X-ray simulation} pipeline, called CT2Xray, which incorporates the differentiable forward projector and produces more realistic radiographs. In~\cite{DBLP:conf/miccai/UnberathZLBFAN18}, a realistic X-ray attenuation $I_{\textrm{atten}}$ is modeled as
\begin{align}
    I_{\textrm{atten}} = \sum_{E}I_{0}e^{-\sum_{m}\mu(m,E)t_{m}} + \textrm{SE} + \textrm{noise},
\end{align}
where $\mu(m,E)$ is the linear attenuation coefficient of material $m$ at energy state $E$ and is measured and known \cite{hubbell1995tables}, $t_{m}$ is the material thickness. $\textrm{SE}$
is the scatter estimation term, $I_{0}$ is the source X-ray intensity. Along with noise, $\textrm{SE}$ and $I_{0}$ are omitted for simplicity.

CT2Xray considers only the bone and tissue materials, that is, $m\in \{\textrm{bone,tissue}\}$. With the aid of a bone mask $V_\textrm{mask}$ and using the differentiable forward projector, it calculates the material thickness wrt the projection parameter $T$ as
\begin{align}
\begin{aligned}
t^{\textrm{CT}}_\textrm{bone}(T)= \textrm{FP}(V_{\textrm{CT}} \odot V_{\textrm{mask}}, T),\\
t^{\textrm{CT}}_\textrm{tissue}(T)= \textrm{FP}(V_{\textrm{CT}} \odot (1-V_{\textrm{mask}}), T).
\end{aligned}
\end{align}
Radiographs are typically stored and viewed as inverted versions of the measured attenuation; therefore, CT2Xray can be expressed as
\begin{align}\label{eq:ct2xray}
    \textrm{CT2Xray}(V_{\textrm{CT}}, V_{\textrm{mask}}, T) = \textrm{max}(I^{T}_{\textrm{atten}}) - I^{T}_{\textrm{atten}},
\end{align}
with
\begin{equation}\label{eq:ct2xray_eq}
I^{T}_{\textrm{atten}} = \sum_{E}e^{-\sum_{m}\mu(m, E)t^{\textrm{CT}}_{m}(T)}.
\end{equation}



\textbf{Single image backprojector (BP)}.
While CT2Xray generates radiographs from CT volumes, an inverse function is needed to transform radiographs back to the respective CT volumes. Such a transformation is clearly ill-posed; however, we can formulate an inverse function of the forward projector to properly place the input X-ray image in 3D. We call such an inverse function the \textit{single image backprojector}. Following a similar formulation from general backprojection algorithm, which reconstructs CBCT from multi-view radiographs as described in \cite{KINAHAN2004421}, the single image backprojector (BP) is expressed as the following:
\begin{equation} \label{eq:bp}
 V^{T}_{\textrm{BP}}(y) = \textrm{BP}(I(x), T) = 
 \begin{cases}
   \frac{I(x)}{\lvert \bm{l}(x) \rvert \Delta p}, &\text{if} ~~ Ty \in \bm{l}(x),\\
     0, &\text{otherwise. }
 \end{cases}
\end{equation}
By substituting $V^{T}_{\textrm{BP}}$ in (\ref{eq:bp}) into $V_{\textrm{CT}}$ in (\ref{eq:drr}), the same $I$ is recovered under view $T$, hence we denote $V^{T}_{\textrm{BP}}$ obtained in this way as the backprojection of $I$ at view $T$. While the same image-wise consistency does not generally apply to CT2Xray, i.e., substituting (\ref{eq:bp}) into (\ref{eq:ct2xray}) does not recover $I$. We show that by using a CNN to complement the single image backprojector, such consistency can be better approximated due to geometric consistency.

\subsection{XraySyn}
XraySyn is trained in two stages, as shown in Fig. \ref{fig:xraysyn_net}. The simulation stage trains a radiograph-to-CT transformation network, called 3D PriorNet, and with help of CT2Xray generate radiographic views from the estimated CT. The real-radiograph stage, which trains a 2D RefineNet, then further closes the domain gap between simulated and real radiographs. Due to the need for calculating the material-dependent attenuation, we also gain the ability to transfer labels from the CT domain to the radiograph domain, in our case with CT bone labels, and achieve bone extraction on real radiographs. 

\textbf{3D PriorNet (3DPN)}.
Under simulated setting, Single Image Backprojector produces $V^{T}_{\textrm{BP}}$ from input view radiograph $I^{T}$, and CT2Xray produces a desired view radiograph $I^{T^{'}}$ from $V_{\textrm{CT}}$ and $V_{\textrm{mask}}$; to complete the end-to-end radiograph-to-CT generation, a function $G$ is needed to recover $V_{\textrm{CT}}$ and $V_{\textrm{mask}}$ from $V^{T}_{\textrm{BP}}$. Mathematically, the generation process between radiograph and CT can be expressed as:
\begin{equation} \label{eq:xraysyn}
\begin{aligned}
    \{V_{\textrm{CT}}, V_{\textrm{mask}}\} = G(\textrm{BP}(I^{T}_{in}, T); \theta), \\ I^{T^{'}}_{\textrm{CT}} = \textrm{CT2Xray}(V_{\textrm{CT}}, V_{\textrm{mask}},T^{'}),
\end{aligned}
\end{equation}
where $\theta$ represents the parameters in $G$. We use a 3D UNet \cite{DBLP:conf/miccai/CicekALBR16} structure for $G$. The loss functions for training $G$ need to ensure consistency both in CT and radiograph domains, and are defined as:
\begin{align}\label{eqn:loss3dpn}
    \mathcal{L}_{G} &= \lambda_{\textrm{CT}}\mathcal{L}_{\textrm{CT}} + \lambda_{\textrm{mask}}\mathcal{L}_{\textrm{mask}} +  \lambda_{\textrm{xray}}(\mathcal{L}^{T}_{\textrm{xray}}+\mathcal{L}^{T^{'}}_{\textrm{xray}}), 
\end{align}
where $\lambda_{\textrm{CT}}$, $\lambda_{\textrm{mask}}$, and $\lambda^{T}_{\textrm{xray}}$ are weights; $\mathcal{L}_{\textrm{CT}}$, $\mathcal{L}_{\textrm{mask}}$, and $\mathcal{L}^{T}_{\textrm{xray}}$ are defined as:
\begin{align}\label{eqn:loss3dpn_2}
\begin{aligned}
\mathcal{L}_{\textrm{CT}} = \lVert V_{\textrm{CT}} - V_{\textrm{gt}} \rVert_1, ~~
\mathcal{L}_{\textrm{mask}} = \textrm{CE}(V_{\textrm{mask}}, V^{\textrm{gt}}_{\textrm{mask}}),\\
\mathcal{L}^{T}_{\textrm{xray}}=\lVert I^{T}_{\textrm{CT}} - I^{T}_{\textrm{in}} \rVert_1 + \sum_{m}\lVert t^{\textrm{CT}}_{\textrm{m}}(T) - t^{\textrm{gt}}_{\textrm{m}}(T) \rVert_1,
\end{aligned}
\end{align}
where CE refers to the cross entropy loss and  $\mathcal{L}^{T^{'}}_{\textrm{xray}}$ is defined similarly as $\mathcal{L}^{T}_{\textrm{xray}}$.


\textbf{2D RefineNet (2DRN)}. While 3DPN $\textrm{G}$ estimates a degree of 3D context from a radiograph, such estimation is both coarse in quality, due to the ill-posed nature, and in resolution, due to memory constraint.
Furthermore, there exists a domain gap between real and simulated radiograph. To address these issues, a second stage, called 2DRN, is introduced. 2DRN has two goals: (i) to generate realistic radiographs from the output of 3DPN with higher resolution, and (ii) to do so with small refinements on $t^{\textrm{CT}}_{m}$ such that we can still obtain the material decomposition of the output radiograph. Conceptually, 2DRN can be understood as a part of an augmented, learnable CT2Xray. 2DRN is constructed in two parts. The main refinement network $\mathcal{F}$ is based on Residual Dense Network (RDN) \cite{DBLP:conf/cvpr/ZhangTKZ018}; additionally, a fully convolutional network $\mathcal{M}$ is used to generate certain convolutional layer parameters in $\mathcal{F}$ directly from the input $I^{T}_{\textrm{in}}$. The purpose of $\mathcal{M}$ is to shuffle high level information that may be lost during the process of 3DPN. Overall, the refinement on $t^{\textrm{CT}}_{m}$ is expressed as:
\begin{align}\label{eqn:loss2drn}
    t^{\textrm{ref}}_{m}(T) =t^{\textrm{CT}}_{m}(T)+ \mathcal{F}(t^{\textrm{CT}}_{m};\phi, 
    \mathcal{M}(I^{T}_{\textrm{in}})), 
\end{align}
where $\phi$ and $\mathcal{M}(I^{T}_{\textrm{in}})$ represent the parameters in $\mathcal{F}$. Replacing $t^{\textrm{CT}}_{m}$ in (\ref{eq:ct2xray}) with $t^{\textrm{ref}}_{m}$ yields an augmented CT2Xray:
\begin{align}\label{eq:ct2xray_aug}
\begin{aligned}
    \textrm{CT2Xray}_{\textrm{aug}}(V_{\textrm{CT}}, V_{\textrm{mask}}, T) = \textrm{max}(I^{T}_{\textrm{ref}}) - I^{T}_{\textrm{ref}},
\end{aligned}
\end{align}
where $I^{T}_{\textrm{ref}} = \sum_{E}e^{-\sum_{m}\mu(m, E)t^{ref}_{m}}$. 
Similarly, the final view synthesis results are defined as $I^{T^{'}}_{\textrm{final}} = \textrm{CT2Xray}_{\textrm{aug}}(V_{\textrm{CT}}, V_{\textrm{mask}},T^{'})$.
A Least-Square GAN (LSGAN) is used to ensure $I^{T^{'}}_{\textrm{final}}$ is statistically similar to $I^{T}_{\textrm{in}}$ when $T^{'}$ and $T$ are relatively close. The overall loss function for 2DRN is described as:
\begin{align}\label{eqn:loss2drn_2}
\mathcal{L}_{\textrm{2DRN}}=\lambda_{\textrm{recon}}\lVert I^{T}_{\textrm{final}} - I^{T}_{\textrm{in}} \rVert_1 + \lambda_{\textrm{GAN}}\mathcal{L}_{\textrm{LSGAN}}(I^{T^{'}}_{\textrm{final}}, I^{T}_{\textrm{in}}).
\end{align}
$\mathcal{L}_{\textrm{LSGAN}}$ is defined as:
\begin{align}\label{eqn:lossGAN}
\begin{aligned}
\mathcal{L}^{\mathcal{D}}_{\textrm{LSGAN}}(I^{T^{'}}_{\textrm{final}}, I^{T}_{\textrm{in}})=\mathbb{E}(\mathcal{D}(I^{T}_{\textrm{im}}-1)^2)) + \mathbb{E}(\mathcal{D}(I^{T^{'}}_{\textrm{final}}-0)^2),\\
\mathcal{L}^{\mathcal{G}}_{\textrm{LSGAN}}(I^{T^{'}}_{\textrm{final}})=\mathbb{E}(\mathcal{D}(I^{T^{'}}_{\textrm{final}}-1)^2),
\end{aligned}
\end{align}
where $\mathcal{G}$ indicates generator, in this case the composition of $G$, $\mathcal{F}$, and $\mathcal{M}$. $\mathcal{D}$ indicates the discriminator. 

\section{Experiments}

\begin{table*}[htb!]
\small
\centering 
\setlength\tabcolsep{1.5pt}
\begin{tabular}{|l|c|c|c|c|c|c|c|l|c|c|}
\hline
\multicolumn{7}{|c|}{Simulated} && \multicolumn{3}{c|}{Real}\\
\hline
\diagbox[innerwidth=6em, height=2em]{Method}{View} & $I^{T}_{\textrm{CT}}$ & $t^{T}_{\textrm{bone}}$ & $t^{T}_{\textrm{tissue}}$ & $I^{T^{'}}_{\textrm{CT}}$ & $t^{T^{'}}_{\textrm{bone}}$ & $t^{T^{'}}_{\textrm{tissue}}$ && \diagbox[innerwidth=6em, height=2em]{Method}{View} & $I^{T}_{\textrm{final}}$ & $I^{T^{'}}_{\textrm{final}}$ (FID) \\
\hline
2D Refiner & \diagbox[innerwidth=5em, height=1.1em]{}{} & \diagbox[innerwidth=5em, height=1.1em]{}{} & \diagbox[innerwidth=5em, height=1.1em]{}{} & 21.09/0.801 & \diagbox[innerwidth=5em, height=1.1em]{}{} & \diagbox[innerwidth=5em, height=1.1em]{}{} & &3DPN & 22.75/0.819  & 1.090\\
\hline
X2CT & 20.44/0.755 & 15.99/0.451 & 19.57/0.711 & 20.20/0.750 & 16.12/0.450 & 19.38/0.703 &&$\textrm{XraySyn}^{\textrm{no}\mathcal{M}}$ &  28.42/0.857  & 0.375\\
\hline
3DPN-DRR & 21.05/0.925 & \diagbox[innerwidth=5em, height=1.1em]{}{} & \diagbox[innerwidth=5em, height=1.1em]{}{} & 19.79/0.867 & \diagbox[innerwidth=5em, height=1.1em]{}{} & \diagbox[innerwidth=5em, height=1.1em]{}{}& &XraySyn & \textbf{30.33/0.865}  & \textbf{0.319} \\
\hline
3DPN & \textbf{29.49/0.961} & \textbf{22.30/0.814} & \textbf{24.40/0.866} & \textbf{27.22/0.929} & \textbf{21.63/0.780} & \textbf{24.27/0.860} && & &\\
\hline

\end{tabular}
\caption{Ablation study of our proposed methods against alternative implementations. Note that 3DPN-DRR and 2D Refiner do not use CT2Xray, therefore are without metrics for $t_{\textrm{bone}}$, $t_{\textrm{tissue}}$. $I^{T}_{\textrm{CT}}$ for 2D Refiner is trivially generated as $I^{T}_{\textrm{in}}$. PSNR/SSIM metrics are provided when groundtruth is available, otherwise FID score is reported. The best performing metrics are \textbf{bold}.} 
\label{tab:ablation_table}

\vspace{-1em}
\end{table*}

\subsubsection{Implementation details}
The two stages of XraySyn are trained separately. To train 3DPN, a CT volume $V_{\textrm{gt}}$ and its bone mask $V^{\textrm{gt}}_{\textrm{mask}}$ is used to simulate the groundtruth radiographs $I^{T}_{\textrm{in}}$ and $I^{T^{'}}_{\textrm{gt}}$. $T$ and $T^{'}$ are sampled randomly from -18$\degree$ to 18$\degree$ in azimuth and elevation angles. During the training of 2DRN, real radiographs are used as $I^{T}_{in}$ in place of simulated radiographs, and the 3DPN's parameters are frozen. Furthermore, the input is first downsampled through average pooling as real radiographs are of higher resolution. As view angles are not available for real radiographs, we sample $T$ and $T^{'}$ randomly in similar fashion as for 3DPN training. For testing on real radiograph, $T$ is assumed to be the canonical frontal view, $T^{'}$ are twenty view angles uniformally spaced from -9$\degree$ to 9$\degree$ in azimuth. Due to the discretization of the voxel-based representation, the ray tracing process used in forward and backprojection needs to approximate points in space when those points are not on the coordinate grid. We use trilinear interpolation for such approximation. The networks are implemented with Pytorch, and trained using four Nvidia P6000 GPUs for five days. The details of network structure are reported in the supplemental material.


\subsubsection{Dataset}
Both CT and radiograph datasets are needed for training XraySyn. To train the 3DPN, we use the LIDC-IDRI dataset \cite{armato2011lung}, which contains
1,018 chest CT volumes. We discards all volumes that have a between-slices resolution higher than 2.5mm. This leads to 780 CT volumes, from which we randomly select 700 for training, 10 for evaluation, and 70 for testing. To preprocess the data, we tightly crop the CT volumes to eliminate excess empty space, and reshape the volumes into resolution of $128\times128\times128$. To train 2DRN with real radiographs, we use the TBX11K dataset \cite{DBLP:conf/cvpr/LiuWBWC20}. Specifically, to avoid the excessive interference of foreign objects and abnormal anatomy, we manually select 3000 images under the healthy category within TBX11K and crop those images to have similar field-of-view as the CT simulation. The images are then reshaped to resolution of $256\times256$. We randomly select 2600 for training, 100 for evaluation, and 300 for testing.






\subsubsection{Evaluation metrics}
Evaluation of novel view synthesis on real radiograph can be challenging, as there is no groundtruth. Therefore, we first provide the results of evaluation on simulated radiographs. Peak Signal-to-Noise Ratio (PSNR) and Structured Similarity Index (SSIM) \cite{DBLP:journals/tip/WangBSS04} are used to measure the quality of synthesized novel views and material decomposition. For real radiographs, we use PSNR and SSIM to measure the quality of reconstruction for the input view $T$ and Fréchet Inception Distance (FID) \cite{DBLP:conf/nips/HeuselRUNH17} to measure the realism of the novel view radiograph in comparison to the input view radiograph. Since radiographs are grayscale images, we do the following to adapt to the ImageNet-trained InceptionV3 model: 1) we repeat the grayscale values across the RGB channels, and 2) use the middle layer features from InceptionV3, specifically the 768-channel layer before InceptionV3’s last auxiliary classifier, for better generalizability.


\captionsetup[subfigure]{labelformat=empty}
\begin{figure*}[htb!]
    \setlength{\abovecaptionskip}{3pt}
    \setlength{\tabcolsep}{1pt}
    \centering
    \begin{tabular}[b]{|ccccccccc|}
        \hline
        View & 2D Refiner & $\textrm{X2CT}_{\textrm{radio}}$ & $\textrm{X2CT}_{\textrm{bone}}$ & 3DPN-DRR & $\textrm{3DPN}_{\textrm{radio}}$ & $\textrm{3DPN}_{\textrm{bone}}$ & $\textrm{GT}_{\textrm{radio}}$ & $\textrm{GT}_{\textrm{bone}}$\\
        
        -9$\degree$ & 
        \begin{subfigure}[b]{0.115\linewidth}
            \includegraphics[width=\textwidth,height=\textwidth]{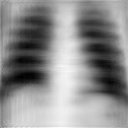}
            \caption{18.85/0.725}
        \end{subfigure} &
        \begin{subfigure}[b]{0.115\linewidth}
            \includegraphics[width=\textwidth,height=\textwidth]{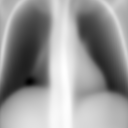}
            \caption{\underline{20.70}/0.745}
        \end{subfigure} &
        \begin{subfigure}[b]{0.115\linewidth}
            \includegraphics[width=\textwidth,height=\textwidth]{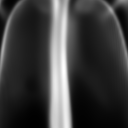}
            \caption{17.07/0.470}
        \end{subfigure} &
        \begin{subfigure}[b]{0.115\linewidth}
            \includegraphics[width=\textwidth,height=\textwidth]{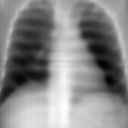}
            \caption{19.28/\underline{0.835}}
        \end{subfigure} &   
        \begin{subfigure}[b]{0.115\linewidth}
            \includegraphics[width=\textwidth,height=\textwidth]{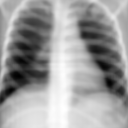}
            \caption{\textbf{26.36/0.918}}
        \end{subfigure} &  
        \begin{subfigure}[b]{0.115\linewidth}
            \includegraphics[width=\textwidth,height=\textwidth]{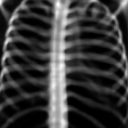}
            \caption{21.47/0.826}
        \end{subfigure} &
        \begin{subfigure}[b]{0.115\linewidth}
            \includegraphics[width=\textwidth,height=\textwidth]{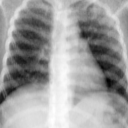}
            \caption{PSNR/SSIM}
        \end{subfigure} &
        \begin{subfigure}[b]{0.115\linewidth}
            \includegraphics[width=\textwidth,height=\textwidth]{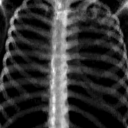}
            \caption{PSNR/SSIM}
        \end{subfigure}\\
        0$\degree$ & 
        \begin{subfigure}[b]{0.115\linewidth}
            \includegraphics[width=\textwidth,height=\textwidth]{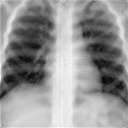}
            \caption{N/A}
        \end{subfigure} &
        \begin{subfigure}[b]{0.115\linewidth}
            \includegraphics[width=\textwidth,height=\textwidth]{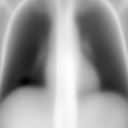}
            \caption{21.52/0.741}
        \end{subfigure} &
        \begin{subfigure}[b]{0.115\linewidth}
            \includegraphics[width=\textwidth,height=\textwidth]{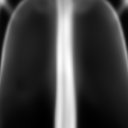}
            \caption{16.83/0.470}
        \end{subfigure} &
        \begin{subfigure}[b]{0.115\linewidth}
            \includegraphics[width=\textwidth,height=\textwidth]{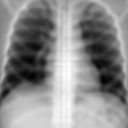}
            \caption{\underline{23.26/0.937}}
        \end{subfigure} &   
        \begin{subfigure}[b]{0.115\linewidth}
            \includegraphics[width=\textwidth,height=\textwidth]{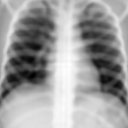}
            \caption{\textbf{31.60/0.971}}
        \end{subfigure} &
        \begin{subfigure}[b]{0.115\linewidth}
            \includegraphics[width=\textwidth,height=\textwidth]{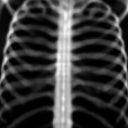}
            \caption{23.67/0.872}
        \end{subfigure} &
        \begin{subfigure}[b]{0.115\linewidth}
            \includegraphics[width=\textwidth,height=\textwidth]{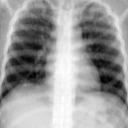}
            \caption{PSNR/SSIM}
        \end{subfigure}&
        \begin{subfigure}[b]{0.115\linewidth}
            \includegraphics[width=\textwidth,height=\textwidth]{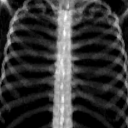}
            \caption{PSNR/SSIM}
        \end{subfigure} \\        
        9$\degree$ & 
        \begin{subfigure}[b]{0.115\linewidth}
            \includegraphics[width=\textwidth,height=\textwidth]{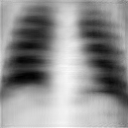}
            \caption{20.11/0.766}
        \end{subfigure} &
        \begin{subfigure}[b]{0.115\linewidth}
            \includegraphics[width=\textwidth,height=\textwidth]{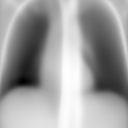}
            \caption{\underline{20.54}/0.724}
        \end{subfigure} &
        \begin{subfigure}[b]{0.115\linewidth}
            \includegraphics[width=\textwidth,height=\textwidth]{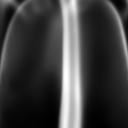}
            \caption{15.72/0.397}
        \end{subfigure} &
        \begin{subfigure}[b]{0.115\linewidth}
            \includegraphics[width=\textwidth,height=\textwidth]{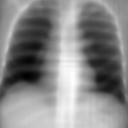}
            \caption{20.07/\underline{0.850}}
        \end{subfigure} &   
        \begin{subfigure}[b]{0.115\linewidth}
            \includegraphics[width=\textwidth,height=\textwidth]{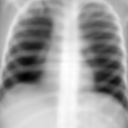}
            \caption{\textbf{27.52/0.927}}
        \end{subfigure} &
        \begin{subfigure}[b]{0.115\linewidth}
            \includegraphics[width=\textwidth,height=\textwidth]{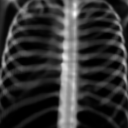}
            \caption{22.44/0.817}
        \end{subfigure} &
        \begin{subfigure}[b]{0.115\linewidth}
            \includegraphics[width=\textwidth,height=\textwidth]{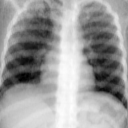}
            \caption{PSNR/SSIM}
        \end{subfigure} &
        \begin{subfigure}[b]{0.115\linewidth}
            \includegraphics[width=\textwidth,height=\textwidth]{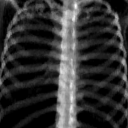}
            \caption{PSNR/SSIM}
        \end{subfigure}\\                
        \hline
    \end{tabular}
    \caption{Visual comparisons of novel view radiographs generated by different methods based on simulated radiographs as inputs. View angle change is measured azimuthally. For each non-groundtruth image, PSNR/SSIM are provided as metrics. Note that the input view for 2D Refiner is trivially generated as $I^{T}_{in}$. Methods that involve the use of CT2Xray are displayed with both the bone $t^{\textrm{CT}}_{\textrm{bone}}$ and radiograph outputs. The best radiograph generation metrics are \textbf{bold}, the second best is \underline{underlined}.}
    \label{tab:ablation_img}
    \vspace{-1.5em}
\end{figure*}
\subsubsection{Ablation study}
We evaluate the effectiveness of XraySyn against alternative 2D or 3D methods. Firstly, comparisons against 3DPN are made under simulated setting for the following setups: 
\begin{itemize}
    \item 2D Refiner: An image-to-image method by synthesizing new views from input $\textrm{FP}(\textrm{BP}(I^{T}_{\textrm{in}}, T),T^{'})$ through a 2D DenseNet structure. The training is constrained by a L1 loss between the generated X-ray and its groundtruth. 
    \item X2CT: Proposed by Ying et al. \cite{DBLP:conf/cvpr/YingGMWWZ19} to transform DRRs into CT volumes. We made the following adaptations: (i) instead of DRRs, the inputs are X-rays generated by CT2Xray, and (ii) the training losses are consistent with Eq. (\ref{eqn:loss3dpn}).
    \item 3DPN-DRR: An alternative 3DPN that directly uses forward projection to simulate radiographs, i.e. DRRs. Note that no material decomposition is involved in DRR.
\end{itemize}
Comparisons against the overall XraySyn method are then made for generating real radiographs. These include:
\begin{itemize}
    \item 3DPN: A direct use of 3DPN trained on simualted data.
    \item $\textrm{XraySyn}^{\textrm{no}\mathcal{M}}$: An alternative XraySyn where the 2DRN stage does not have $\mathcal{M}$.
\end{itemize}


Table \ref{tab:ablation_table} summarizes the performance of different implementations over various view angles, and visualization of the results are provided in Fig. \ref{tab:ablation_img}. Image-to-image approach like 2D Refiner synthesizes mostly a blurry version of the input view. The lack of an explicit 3D loss forces the network to overly smooth the output for the best PSNR. The implementation of X2CT-CNN copies the 2D features along a third axis to achieve the initial upsampling from 2D to 3D, which is geometrically incorrect for arbitrary view angle. While CNN's strong learning ability still helps X2CT-CNN produce a coarse 3D estimation, it has difficulty in learning the geometric transformation that maps the input image to the correct view. Consequently, the synthesized views are reliant on a poorly estimated 3D structure and lack significant details, specifically over the rib cage area. 
\captionsetup[subfigure]{labelformat=empty}
\begin{figure*}[htb!]
    \setlength{\abovecaptionskip}{3pt}
    \setlength{\tabcolsep}{1pt}
    \centering
    \begin{tabular}[b]{|cccccccc|}
        \hline
        View & $\textrm{3DPN}_{\textrm{radio}}$ & $\textrm{3DPN}_{\textrm{bone}}$ & $\textrm{XraySyn}^{\textrm{no}\mathcal{M}}_{\textrm{radio}}$ & $\textrm{XraySyn}^{\textrm{no}\mathcal{M}}_{\textrm{bone}}$ & $\textrm{XraySyn}_{\textrm{radio}}$ & $\textrm{XraySyn}_{\textrm{bone}}$ & GT/Input\\
        
        -9$\degree$ & 
        \begin{subfigure}[b]{0.125\linewidth}
            \includegraphics[width=\textwidth,height=\textwidth]{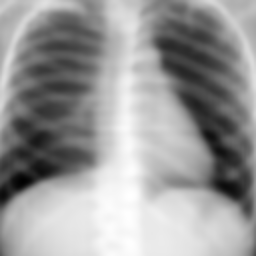}
            \caption{0.903}
        \end{subfigure} &
        \begin{subfigure}[b]{0.125\linewidth}
            \includegraphics[width=\textwidth,height=\textwidth]{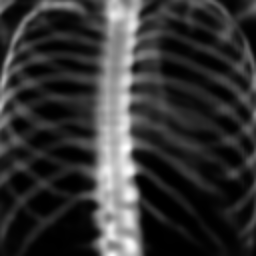}
            \caption{-/-}
        \end{subfigure} &
        \begin{subfigure}[b]{0.125\linewidth}
            \includegraphics[width=\textwidth,height=\textwidth]{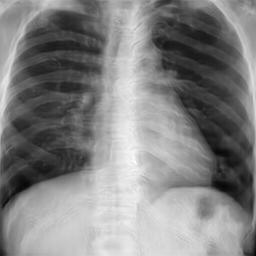}
            \caption{0.414}
        \end{subfigure} &
        \begin{subfigure}[b]{0.125\linewidth}
            \includegraphics[width=\textwidth,height=\textwidth]{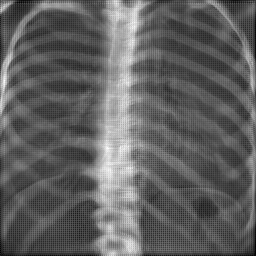}
            \caption{-/-}
        \end{subfigure} &   
        \begin{subfigure}[b]{0.125\linewidth}
            \includegraphics[width=\textwidth,height=\textwidth]{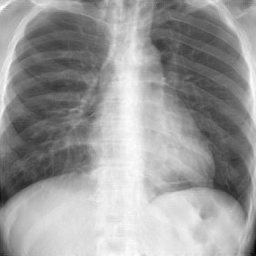}
            \caption{\textbf{0.368}}
        \end{subfigure} &  
        \begin{subfigure}[b]{0.125\linewidth}
            \includegraphics[width=\textwidth,height=\textwidth]{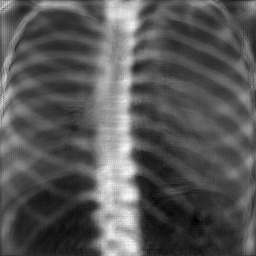}
            \caption{-/-}
        \end{subfigure} &
        \begin{subfigure}[b]{0.125\linewidth}
            \includegraphics[width=\textwidth,height=\textwidth]{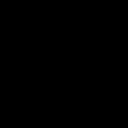}
            \caption{FID}
        \end{subfigure} \\
        0$\degree$ & 
        \begin{subfigure}[b]{0.125\linewidth}
            \includegraphics[width=\textwidth,height=\textwidth]{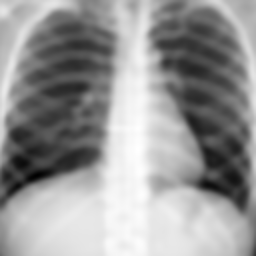}
            \caption{23.18/0.832}
        \end{subfigure} &
        \begin{subfigure}[b]{0.125\linewidth}
            \includegraphics[width=\textwidth,height=\textwidth]{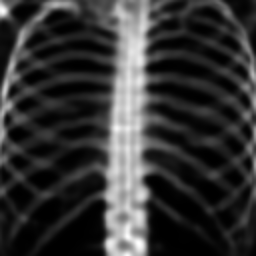}
            \caption{N/A}
        \end{subfigure} &
        \begin{subfigure}[b]{0.125\linewidth}
            \includegraphics[width=\textwidth,height=\textwidth]{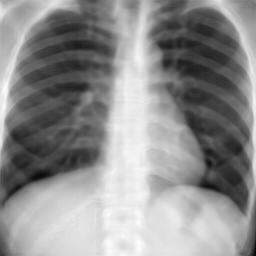}
            \caption{24.22/0.845}
        \end{subfigure} &
        \begin{subfigure}[b]{0.125\linewidth}
            \includegraphics[width=\textwidth,height=\textwidth]{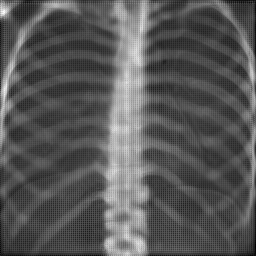}
            \caption{N/A}
        \end{subfigure} &   
        \begin{subfigure}[b]{0.125\linewidth}
            \includegraphics[width=\textwidth,height=\textwidth]{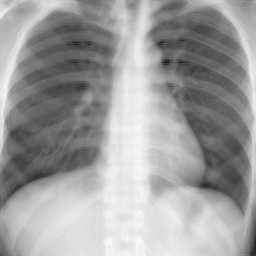}
            \caption{\textbf{30.60/0.861}}
        \end{subfigure} &  
        \begin{subfigure}[b]{0.125\linewidth}
            \includegraphics[width=\textwidth,height=\textwidth]{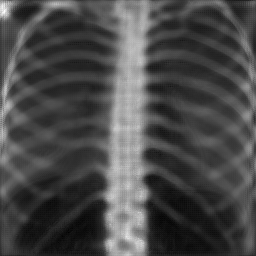}
            \caption{N/A}
        \end{subfigure} &
        \begin{subfigure}[b]{0.125\linewidth}
            \includegraphics[width=\textwidth,height=\textwidth]{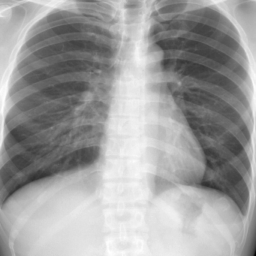}
            \caption{PSNR/SSIM}
        \end{subfigure} \\        
        9$\degree$ & 
        \begin{subfigure}[b]{0.125\linewidth}
            \includegraphics[width=\textwidth,height=\textwidth]{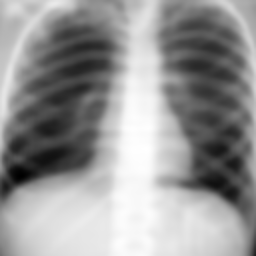}
            \caption{0.929}
        \end{subfigure} &
        \begin{subfigure}[b]{0.125\linewidth}
            \includegraphics[width=\textwidth,height=\textwidth]{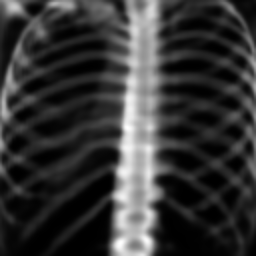}
            \caption{-/-}
        \end{subfigure} &
        \begin{subfigure}[b]{0.125\linewidth}
            \includegraphics[width=\textwidth,height=\textwidth]{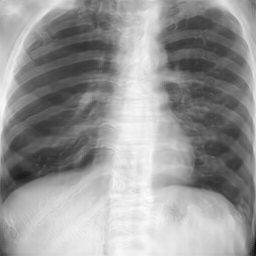}
            \caption{0.432}
        \end{subfigure} &
        \begin{subfigure}[b]{0.125\linewidth}
            \includegraphics[width=\textwidth,height=\textwidth]{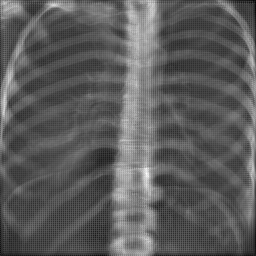}
            \caption{-/-}
        \end{subfigure} &   
        \begin{subfigure}[b]{0.125\linewidth}
            \includegraphics[width=\textwidth,height=\textwidth]{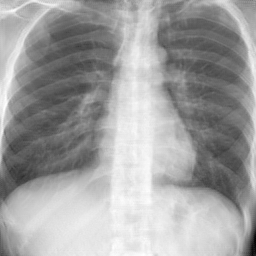}
            \caption{\textbf{0.414}}
        \end{subfigure} &  
        \begin{subfigure}[b]{0.125\linewidth}
            \includegraphics[width=\textwidth,height=\textwidth]{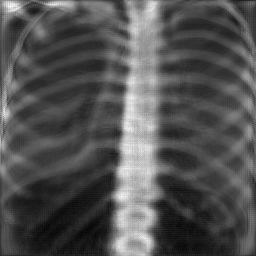}
            \caption{-/-}
        \end{subfigure} &
        \begin{subfigure}[b]{0.125\linewidth}
            \includegraphics[width=\textwidth,height=\textwidth]{imgs/ablation/blank.png}
            \caption{FID}
        \end{subfigure}\\                
        \hline
    \end{tabular}
    \caption{Visual comparisons of novel view radiographs generated by different methods based on real radiographs as inputs. View angle change is measured azimuthally. PSNR/SSIM metrics are provided for input radiograph reconstruction, FID is provided for novel view synthesis with respect to input radiographs. The best metrics are \textbf{bold}. For more results with animation, please refer to the supplemental material.}
    \label{tab:ablation_img_real}
    \vspace{-1.5em}
\end{figure*}
By using the geometrically consistent backprojection and forward projection, 3DPN-DRR and 3DPN perform much better at preserving the input radiograph during the 2D-to-3D transformation. The main issue of DRR is its assumption that the rays attenuate over bone and tissue similarly, when in fact bone attenuates X-Rays much better and therefore has better contrast on real radiographs. The results show that when 3DPN-DRR is used on more realistically simulated radiographs, bone appears much softer. While this impacts PSNR significantly, 3DPN-DRR results have much better SSIM scores compared to X2CT and 2D Refiner due to the superior 3D estimation. Finally, 3DPN, which includes all proposed components, performs much better than other methods (6-7dB more than the second best method in terms of PSNR) at capturing the 3D anatomy from a single radiograph. As 2D-to-3D transformation is still ill-posed, the metrics worsen when the novel view is further from the input view. Please see the supplementary material section for more detailed metrics.

Using 3DPN to perform view synthesis on real radiographs involves additional challenges. Specifically, 3DPN is limited in the available resolution and cannot generate sufficient details, e.g. as shown in Fig. \ref{tab:ablation_img_real}, 3DPN-generated bone structure is less visible on the rib cage area, as those bones are small and hard to be accurately estimated in 3D. Furthermore, simulation does not address complex imaging conditions. As a result, views synthesized through 3DPN are blurry and constrained to have a certain image style due to the formulation of CT2Xray. 2DRN is proposed to address these problems by performing refinement on $t^{\textrm{CT}}_{m}(T^{'})$ to preserve the material decomposition information. 

After refining the 3DPN results in $\textrm{XraySyn}^{\textrm{no}\mathcal{M}}$, the novel views are much improved in both low-level details and overall realism, along with the corresponding bone map $t^{\textrm{ref}}_{\textrm{bone}}$. However, some information is inevitably lost during 3DPN. The input in Fig. \ref{tab:ablation_img_real} is grayish in the lung area, as opposed to the typical dark color from simulation. This information is not captured in 3DPN, thus the subsequent refinement network is also limited in its recovery performance. $\mathcal{M}$ is designed to shuffle information directly from the input radiograph. To prevent 2DRN from learning an identity transform of the input radiograph, $\mathcal{M}$ generates convolutional filter parameters to constrict the information flow. The complete XraySyn, which includes $\mathcal{M}$ in 2DRN, improves the overall performance \begin{figure}[!htb]
    \setlength{\abovecaptionskip}{3pt}
    \setlength{\tabcolsep}{2pt}
    \begin{tabular}[b]{cccc}
        \begin{subfigure}[b]{.24\linewidth}
            \includegraphics[width=\textwidth,height=\textwidth]{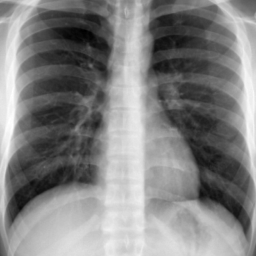}
        \end{subfigure} &
        \begin{subfigure}[b]{.24\linewidth}
            \includegraphics[width=\textwidth,height=1\textwidth]{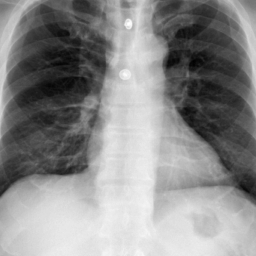}
        \end{subfigure} 
        \begin{subfigure}[b]{.24\linewidth}
            \includegraphics[width=\textwidth,height=1\textwidth]{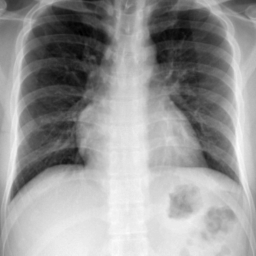}
        \end{subfigure} &
        \begin{subfigure}[b]{.24\linewidth}
            \includegraphics[width=\textwidth,height=1\textwidth]{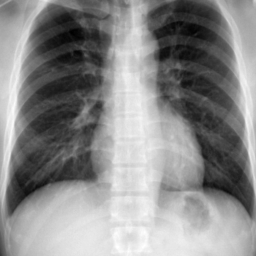}
        \end{subfigure} \\
        \begin{subfigure}[b]{.24\linewidth}
            \includegraphics[width=\textwidth,height=\textwidth]{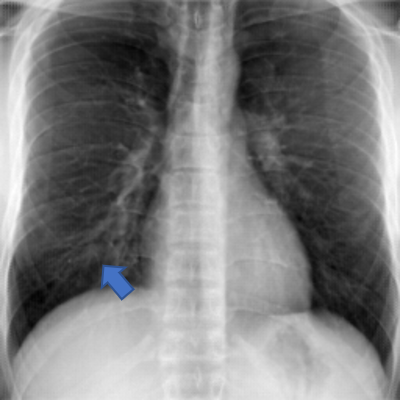}
        \end{subfigure} &
        \begin{subfigure}[b]{.24\linewidth}
            \includegraphics[width=\textwidth,height=1\textwidth]{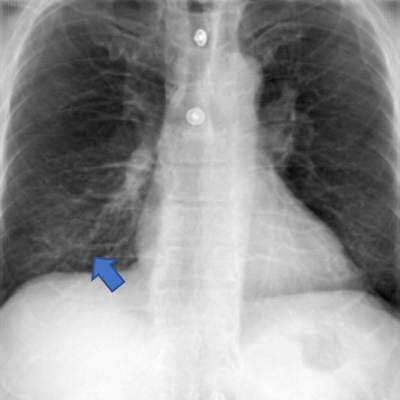}
        \end{subfigure} 
        \begin{subfigure}[b]{.24\linewidth}
            \includegraphics[width=\textwidth,height=1\textwidth]{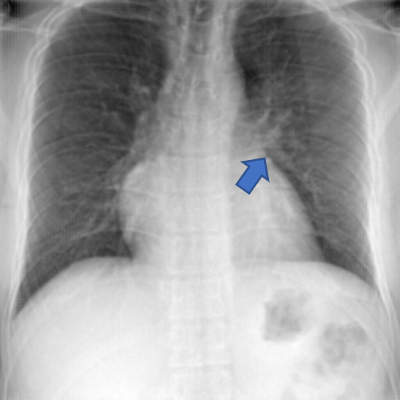}
        \end{subfigure} &
        \begin{subfigure}[b]{.24\linewidth}
            \includegraphics[width=\textwidth,height=1\textwidth]{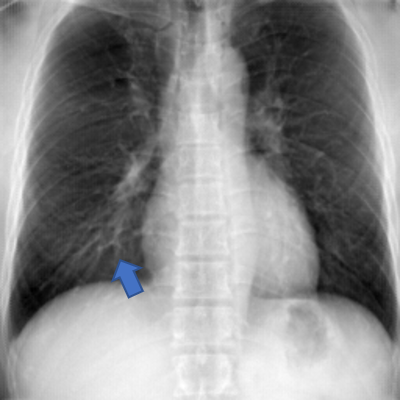}
        \end{subfigure} \\    
    \end{tabular}
    \caption{Bone-suppressed radiographs obtained by post-processing the results from XraySyn. (top) Input radiographs from TBX11K \cite{DBLP:conf/cvpr/LiuWBWC20} and (bottom) the bone-suppressed results. Note that the arteries, as indicated with blue arrows, are more visible after bone suppression.}
    \label{fig:suppress}
    \vspace{-2em}
\end{figure}by almost 1dB with little additional computational cost. While 2DRN does not explicitly guarantee overall 3D consistency, we observe that the novel views are fairly consistent with each other due to the residual learning on the 3D-consistent 3DPN results. As the refinement on $t^{\textrm{CT}}_{m}$ is not directly supervised, we observe a slight amount of background noise despite training the network with a residual connection. However, this can be addressed through post-processing steps. It is worth noting that due to the lack of alternative-view radiograph dataset, synthesized novel views beyond a limited range from the frontal view may lead to unfaithful results as compared to real circumstances.




\subsubsection{Bone suppression}
A natural application of XraySyn is on radiograph bone suppression, which seeks to reduce bone attenuation and better reveal the underlying tissues. To best preserve the tissue information from input radiograph $I^{T}_{\textrm{in}}$, we find $t^{\textrm{ref}}_{\textrm{bone}}$ and $t^{\textrm{recon}}_{\textrm{tissue}}$ so that they losslessly reconstruct $I^{T}_{\textrm{in}}$, through reversing the CT2Xray operation as shown in Eq. (\ref{eq:ct2xray_eq}). In Fig. \ref{fig:suppress}, we show that our approach suppresses most of the rib cage bones while preserving the tissue content. For a mathematical formulation on reversing CT2Xray, please refer to the supplemental material section.

\section{Conclusion}
We propose a two-stage radiograph view synthesis method, XraySyn. This method estimates a coarse 3D CT from a 2D radiograph, simulates a novel view from the estimated volume, and finally refines the views to be visually consistent with real radiographs. The learning process of XraySyn is enabled by our proposed differentiable forward projector and backprojector. Furthermore, by incorporating the CT bone labels in CT2Xray that is inspired by DeepDRR and implemented with our differentiable forward projector, we not only achieve realistic simulation for training the radiograph-to-CT transformation, but also gain the ability to transfer bone labels from CT to radiograph. We carefully evaluate our method both on simulated and real radiographs, and find that XraySyn generates highly realistic and consistent novel view radiographs. To the best of our knowledge, this is the first work on radiograph view synthesis, which can help give practitioners a more precise understanding of the patient's 3D anatomy. XraySyn also opens up possibilities for many downstream processes on radiograph, such as lesion detection, organ segmentation, and sparse-view CT reconstruction. Currently, XraySyn is limited in resolution due to memory constraint of 3DPN. We plan to address this limitation through a more efficient network design. In addition, we will conduct a study to assess the effect of bone suppression for clinical diagnosis.



{\small
\bibliography{main}
}

\newpage
\appendix

\begin{figure*}[]
\centering
\includegraphics[width=\textwidth]{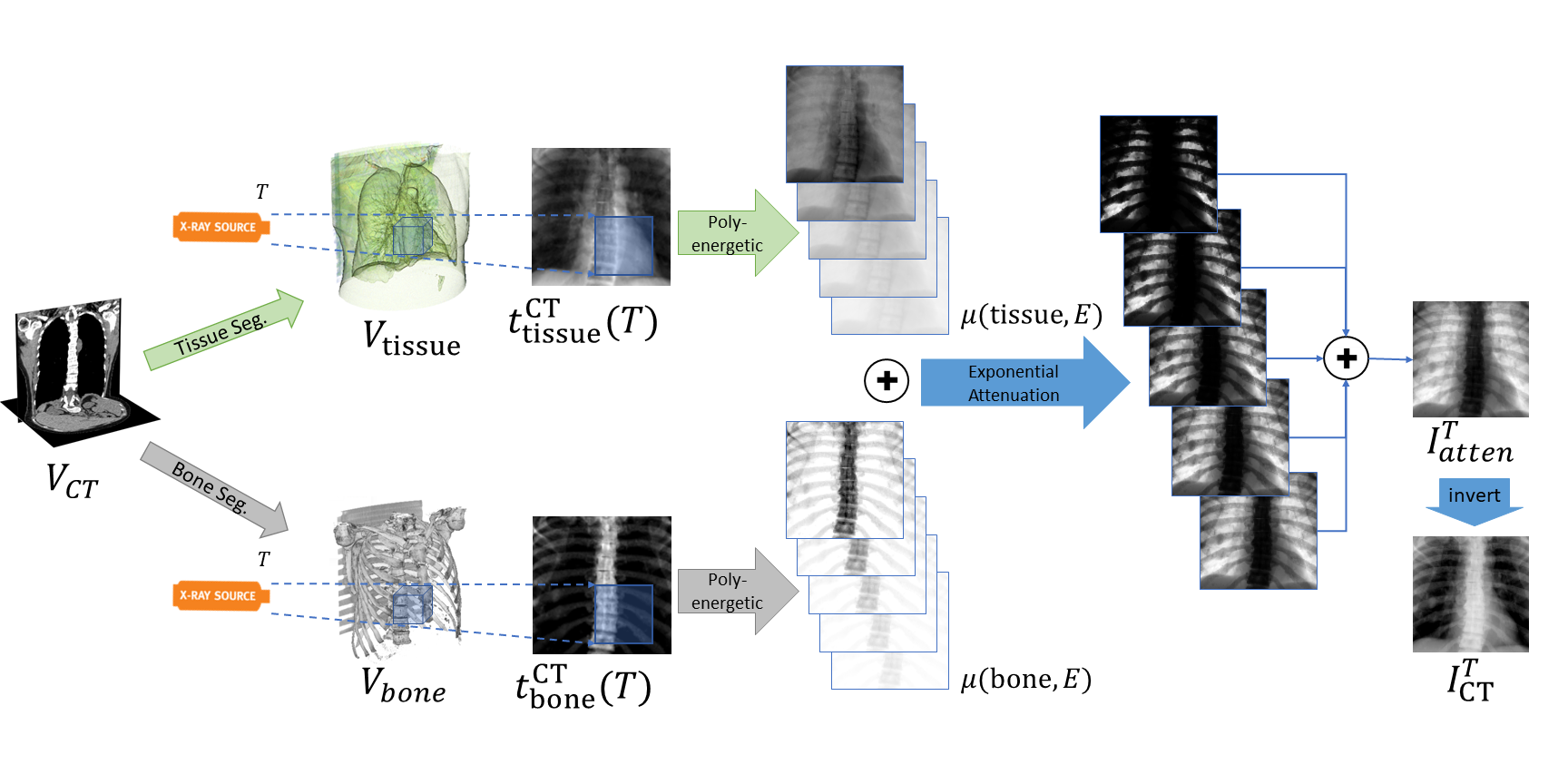}
\caption{A visualization of the CT2Xray component within XraySyn.}
\label{fig:ct2xray_vis}
\end{figure*}
\section{Network architecture}

The proposed XraySyn consists of 3D PriorNet (3DPN) and 2D RefineNet (2DRN). The general model architectures of 3DPN and 2DRN follow 3D-UNet \cite{DBLP:conf/miccai/CicekALBR16} and RDN \cite{DBLP:conf/cvpr/ZhangTKZ018}. For 3DPN, the exact network structure is presented in Table \ref{table:3DPN-fg}. For 2DRN, we use RDN with sixteen Residual Dense Blocks (RDBs), eight convolutional layers per RDB, and growth rate of sixty-four. After the Global Feature Fusion (GFF) layer, an additional convolutional layer with kernel size of one is introduced where the filter parameters are generated by a subnetwork $\mathcal{M}$. The upsampling layers then upsample spatial dimensions from $128 \times 128$ to $256 \times 256$, the output channel dimension is two, for bone and tissue. The exact network structure of $\mathcal{M}$ is presented in Table \ref{table:2DRN-M-fg} and is a fully convolutional network. As the spatial dimension decreases four times at every convolutional layer with a stride of two, $\mathcal{M}$ is computationally efficient both in memory and speed. In the network tables, $N_c$ denotes the number of output channels, $C'$ denotes the channel dimension of generated features $F^{LR}$. We use `K\#-C\#-S\#-P\#' to denote the configuration of the convolution layers, where `K', `C', `S' and `P' stand for the kernel, input channel, stride and padding size, respectively. 
\begin{table}[htb!]
\centering 
\begin{tabular}{l | c | l}
Name & $N_{c}$ & Description \\
\hline
INPUT & $1$ & Input $I^{T}_{\textrm{in}}$\\ 
CONV0 & $32$ & K3-C1-S2-P1 \\
RELU &  & \\
CONV1 & $64$ & K3-C64-S2-P1 \\
RELU &  & \\
CONV2 & $128$ & K3-C64-S2-P1 \\
RELU &  & \\
CONV3 & $256$ & K3-C128-S2-P1 \\
RELU &  & \\
CONV4 & $512$ & K3-C256-S2-P1 \\
RELU &  & \\
CONV5 & $1024$ & K3-C512-S2-P1 \\
RELU &  & \\
CONV6 & $2048$ & K3-C1024-S2-P1 \\
RELU &  & \\
CONV4 & $4096$ & K3-C2048-S2-P1 \\
\midrule
\end{tabular}
\caption{Network architecture of $\mathcal{M}$ in 2DRN.}
\label{table:2DRN-M-fg}
\end{table}
\begin{table}[htb!]
\centering 
\begin{tabular}{l | c | l}
Name & $N_{c}$ & Description \\
\hline
INPUT & $1$ & Input $V_{BP}$\\ 
CONV0 & $96$ & K3-C1-S2-P1 \\
LEAKY RELU &  & Negative\_slope=0.2\\

CONV1 & $192$ & K3-C96-S2-P1 \\
BN & & BatchNorm3D \\
LEAKY RELU &  & Negative\_slope=0.2\\

CONV2 & $384$ & K3-C192-S2-P1 \\
BN & & BatchNorm3D\\
LEAKY RELU &  & Negative\_slope=0.2\\

CONV3 & $768$ & K3-C384-S2-P1 \\
BN & & BatchNorm3D\\
LEAKY RELU &  & Negative\_slope=0.2\\

CONV4 & $768$ & K3-C768-S2-P1 \\
BN & & BatchNorm3D\\
LEAKY RELU &  & Negative\_slope=0.2\\

CONV5 & $768$ & K3-C768-S2-P1 \\
BN & & BatchNorm3D\\
LEAKY RELU &  & Negative\_slope=0.2\\

UPX2 & & Trilinear Interpolation\\
CONV6 & $768$ & K3-C768-S1-P1 \\
BN & & BatchNorm3D\\
RELU &  & \\

UPX2 & & Trilinear Interpolation\\
CONV7 & $768$ & K3-C1536-S1-P1 \\
BN & & BatchNorm3D\\
RELU &  & \\

UPX2 & & Trilinear Interpolation\\
CONV8 & $384$ & K3-C1536-S1-P1 \\
BN & & BatchNorm3D\\
RELU &  & \\

UPX2 & & Trilinear Interpolation\\
CONV9 & $192$ & K3-C768-S1-P1 \\
BN & & BatchNorm3D\\
RELU &  & \\

UPX2 & & Trilinear Interpolation\\
CONV10 & $96$ & K3-C384-S1-P1 \\
BN & & BatchNorm3D\\
RELU &  & \\

UPX2 & & Trilinear Interpolation\\
CONV10 & $2$ & K3-C192-S1-P1 \\
TANH & & \\
\midrule
\end{tabular}
\caption{Network architecture of 3DPN. BN stands for 3D BatchNorm, and UPX2 stands for an upsampling operation by using trilinear interpolation. All convolutional layers are 3D convolutions.}
\label{table:3DPN-fg}
\end{table}

2D Refiner as described in Ablation study also uses the RDN structure for refinement on $\textrm{FP}(\textrm{BP}(I^{T}_{\textrm{in}}, T),T^{'})$, with twenty Residual Dense Blocks (RDBs), eight convolutional layers per RDB, and growth rate of sixty-four. The input for 2D Refiner is a concatenation of $\textrm{FP}(\textrm{BP}(I^{T}_{\textrm{in}}, T),T^{'})$ and $I^{T}_{\textrm{in}}$ to ensure both the original information and the geometric transformation are included.

\begin{algorithm*}[!htb]
\SetAlgoLined
\KwIn{input radiograph: $I^{T}_{\textrm{in}}$, estimated bone and tissue: $t^{\textrm{ref}}_{\textrm{bone}}(T)$ and $t^{\textrm{ref}}_{\textrm{tissue}}(T)$, estimated radiograph attenuation: $I^{T}_{\textrm{atten}}$}, attenuation coefficient: $\mu(m, E)$

\KwOut{reconstructed tissue: $t^{\textrm{recon}}_{\textrm{tissue}}$, bone suppressed radiograph: $I^{T}_{\textrm{inSuppress}}$}

$I^{T}_{\textrm{inAtten}} = \frac{\textrm{max}(I^{T}_{\textrm{in}})-I^{T}_{\textrm{in}}}{\textrm{max}(I^{T}_{\textrm{in}})}$ \tcp*{invert and normalize $I^{T}_{\textrm{in}}$}

$I^{T}_{\textrm{inAtten}} = I^{T}_{\textrm{inAtten}}*(\textrm{max}(I^{T}_{\textrm{atten}})-\textrm{min}(I^{T}_{\textrm{atten}})) + \textrm{min}(I^{T}_{\textrm{atten}}))$ \tcp*{normalize $I^{T}_{\textrm{in}}$ to the range of $I^{T}_{\textrm{atten}}$}

Given $I^{T}_{\textrm{atten}} = \sum_{E}e^{-\sum_{m}\mu(m, E)t^{\textrm{ref}}_{m}(T)} = e^{-\sum_{m}\mu(m, E_{1})t^{\textrm{ref}}_{m}(T)} + \cdots + e^{-\sum_{m}\mu(m, E_{N})t^{\textrm{ref}}_{m}(T)} = w_{1}*I^{T}_{\textrm{atten}} +  \cdots + w_{N}*I^{T}_{\textrm{atten}}$

$w_{i} = \frac{e^{-\sum_{m}\mu(m, E_{i})t^{\textrm{ref}}_{m}(T)}}{I^{T}_{atten}}$ \tcp{estimate decomposition weights to reverse summation $\sum_{E}$}

Given $w_{i}*I^{T}_{\textrm{inAtten}} = e^{-(\mu(\textrm{bone}, E_{i})t^{\textrm{ref}}_{\textrm{bone}}(T) + \mu(\textrm{tissue}, E_{i})t^{\textrm{recon}}_{\textrm{tissue}}(T))}$

$t^{\textrm{recon}}_{\textrm{tissue}}(T, i) = \frac{\textrm{log}(w_{i}*I^{T}_{\textrm{inAtten}}) - \mu(\textrm{bone}, E_{i})t^{\textrm{ref}}_{\textrm{bone}}(T)}{\mu(\textrm{tissue}, E_{i})}$ \tcp*{$t^{\textrm{recon}}_{\textrm{tissue}}(T, i)$ is slightly different based on $w_{i}$}

$I^{T}_{\textrm{inSuppress}} = \sum_{E}e^{-\mu(\textrm{tissue}, E)(t^{\textrm{ref}}_{\textrm{bone}}(T) + t^{\textrm{recon}}_{\textrm{tissue}}(T, E))}$ \tcp{replace the attenuation coeff. of bone to tissue}

 \caption{Reverse CT2Xray to achieve bone suppression}
 \label{alg:reverse_CT2Xray}
\end{algorithm*}

\section{CT2Xray}
A more detailed visualization of CT2Xray is shown in Fig. \ref{fig:ct2xray_vis}. Given the CT volume and its bone label $\{V_{\textrm{CT}}, V_{\textrm{bone}}\}$, we can calculate for $\{V_{bone}, V_{tissue}\}$ and obtain $t^{\textrm{CT}}_{\textrm{bone}}(T), t^{\textrm{CT}}_{\textrm{tissue}}(T)$ through our proposed differential forward projection. The corresponding attenuation is then calculated and summed to produce a realistic radiograph. Compared to the original DeepDRR \cite{DBLP:conf/miccai/UnberathZLBFAN18}, CT2Xray discarded calculation for air-specific attenuation (air volume is considered as tissue), scatter estimation, and noise.
We find that air-specific attenuation calculation as followed by the measurements in \cite{hubbell1995tables} leads to a reduction in contrast on our CT dataset, and that the original CNN-based scatter estimation in DeepDRR is not robust enough on unseen data. 
\begin{table*}[htb!]
\small
\centering 
\setlength\tabcolsep{1.5pt}
\begin{tabular}{|l|c|c|c|c|c|c|c|c|c|c|}
\hline

\diagbox[innerwidth=6em, height=2em]{Method}{View} & -9\degree & -7\degree & -5\degree & -3\degree & -1\degree & 1\degree & 3\degree & 5\degree & 7\degree & 9\degree \\
\hline
2D Refiner &19.29/0.754 &20.32/0.774 &21.23/0.800 &21.98/0.832 &22.40/0.858 & 22.38/0.855& 21.84/0.830& 21.06/0.801& 20.30/0.778&19.56/0.761\\
\hline
X2CT & 19.46/0.738 & 19.93/0.744 & 20.18/0.748 & 20.30/0.751 & 20.41/0.753 & 20.42/0.755 & 20.39/0.756 & 2031/0.754 & 20.18/0.752 & 19.98/0.747\\
\hline
3DPN-DRR & 18.54/0.817& 19.33/0.837& 19.97/0.859& 20.55/0.883& 21.18/0.910& 20.60/0.909&20.10/0.884 & 19.37/0.860&18.63/0.839 &18.03/0.819\\
\hline
3DPN & \textbf{25.37/0.904} & \textbf{25.91/0.913} & \textbf{26.60/0.923} & \textbf{27.55/0.935} & \textbf{28.78/0.951} & \textbf{28.91/0.951} & \textbf{27.93/0.937} & \textbf{27.01/0.926} & \textbf{26.30/0.916}&\textbf{25.74/0.907}\\
\hline

\end{tabular}
\caption{Ablation study of 3DPN against alternative implementations on simulated data over synthesized radiographic views from -9\degree to 9\degree. PSNR/SSIM metrics are provided. The best performing metrics are \textbf{bold}.} 
\label{tab:ablation_table_supp}

\end{table*}
\section{Reversing CT2Xray for bone suppression}
\begin{figure}[htb!]
    \setlength{\abovecaptionskip}{3pt}
    \setlength{\tabcolsep}{1pt}
    \begin{tabular}[b]{ccc}
        \begin{subfigure}[b]{0.33\linewidth}
            \includegraphics[width=\textwidth]{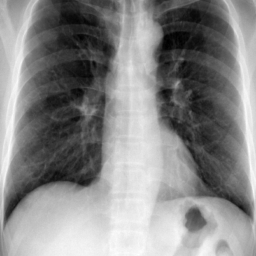}
            \caption{Input $I^{T}_{\textrm{in}}$}
        \end{subfigure} &
        \begin{subfigure}[b]{0.33\linewidth}
            \includegraphics[width=\textwidth]{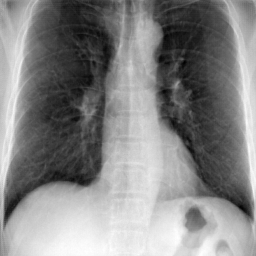}
            \caption{$\{t^{\textrm{ref}}_{\textrm{bone}}, t^{\textrm{recon}}_{\textrm{tissue}}\}$}
        \end{subfigure}&
        \begin{subfigure}[b]{0.33\linewidth}
            \includegraphics[width=\textwidth]{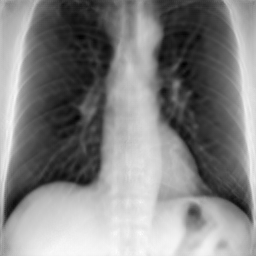}
            \caption{$\{t^{\textrm{ref}}_{\textrm{bone}}, t^{\textrm{ref}}_{\textrm{tissue}}\}$}
        \end{subfigure} \\
    \end{tabular}
    \caption{Visual comparison of bone suppression generated from $\{t^{\textrm{ref}}_{\textrm{bone}}, t^{\textrm{recon}}_{\textrm{tissue}}\}$ and $\{t^{\textrm{ref}}_{\textrm{bone}}, t^{\textrm{ref}}_{\textrm{tissue}}\}$. Note that details are much clearer in the $\{t^{\textrm{ref}}_{\textrm{bone}}, t^{\textrm{recon}}_{\textrm{tissue}}\}$ generated image.}
    \label{fig:bone_sup_supp}
\end{figure}

Bone suppression seeks to reduce bone attenuation such that the overlapped tissue is more visible. While $t^{\textrm{ref}}_{\textrm{bone}}$ and $t^{\textrm{ref}}_{\textrm{tissue}}$ are generated from XraySyn and can be used straightforwardly to reduce bone attenuation, they do not preserve all the details in input $I^{T}_{\textrm{in}}$. Therefore, we seek to reverse CT2Xray mathematically to find a $t^{\textrm{recon}}_{\textrm{tissue}}$, such that $t^{\textrm{ref}}_{\textrm{bone}}$ and $t^{\textrm{recon}}_{\textrm{tissue}}$ reconstruct $I^{T}_{\textrm{in}}$ losslessly. As shown in Algorithm \ref{alg:reverse_CT2Xray}, we first transform $I^{T}_{\textrm{in}}$ from the range of image pixels to physical attenuation by using $I^{T}_{\textrm{atten}}$ (described in Eq. \ref{eq:ct2xray_eq}) as the reference, and obtain $I^{T}_{\textrm{inAtten}}$. We then use $t^{\textrm{ref}}_{\textrm{tissue}}$ as a reference to calculate the weight of each term in the summation over energy level $E$ that produced $I^{T}_{\textrm{atten}}$, and apply these weights on $I^{T}_{\textrm{inAtten}}$. $t^{\textrm{recon}}_{\textrm{tissue}}$ can then be calculated straightforwardly and used in the forward attenuation calculations to obtain bone-suppressed radiograph $I^{T}_{\textrm{inSupress}}$, where the attenuation coefficient of $t^{\textrm{ref}}_{\textrm{bone}}$ is reduced. A comparison of bone suppression by using $\{t^{\textrm{ref}}_{\textrm{bone}}, t^{\textrm{ref}}_{\textrm{tissue}}\}$ and $\{t^{\textrm{ref}}_{\textrm{bone}}, t^{\textrm{recon}}_{\textrm{tissue}}\}$ are shown in Fig. \ref{fig:bone_sup_supp}. 

\captionsetup[subfigure]{labelformat=empty}

\section{More detailed metrics on simualted data}
A more detailed view synthesis evaluation is shown in Table \ref{tab:ablation_table_supp}, where PSNR/SSIM is measured over ten different synthesized views. As we can see, view synthesis becomes less accurate with respect to the ground truth as the view angle deviates further away from the input view. This is consistent with intuition as a single view contains very limited 3D information.

\section{Additional visual results}
For additional visual results of XraySyn on radiograph view synthesis and bone suppression, please refer to the attached multi-media file.

\end{document}